\documentclass[
    ,final            % use final for the camera ready runs
  ]
  {aipproc}

\layoutstyle{6x9}

\begin{document}

\title{The Future of Microwave Background Physics}

\author{Arthur Kosowsky}{
  address={Department of Physics and Astronomy, Rutgers University, 
           136 Frelinghuysen Road, Piscataway, NJ 08854-8019}
}

\begin{abstract}
The cosmic microwave background is now fulfilling its promise of determining
the basic cosmological parameters describing our Universe.  Future
study of the microwave background will mostly be directed towards two
basic questions: a complete characterization of the initial
perturbations, and probes of the nonlinear evolution of structure in
the Universe. The basic scientific issues in both of these areas are
reviewed here, along with possibilities for addressing them with further
microwave background measurements at higher sensitivities and smaller
angular scales. The proposed ACT experiment, which will map 200
square degrees of sky at arcminute resolution and micro-Kelvin sensitivity
in three microwave frequency bands, is briefly described as an example
of rapidly advancing experimental technique.
\end{abstract}

\maketitle

\section{Where We Are Now}

The cosmic microwave background is one of the best-studied
sources of cosmological information (see \cite{kk99,dur01,hd02} for reviews). 
It is nearly isotropic
on the sky, with small temperature fluctuations at the level
of one part in $10^5$. These fluctuations arise from several
basic physical mechanisms in the early Universe at a redshift
$z\simeq 1100$: gravitational redshift, temperature
fluctuations at the last scattering surface, Doppler shifts
from peculiar velocities at the last scattering surface, and
diffusion damping through the thickness of the last scattering
surface \cite{kos01,hu03}. 
In addition, further temperature fluctuations and spectral
distortions arise from gravitational and scattering effects
at comparatively recent epochs, with $z < 3$. Simple arguments
show that the microwave background should also have small polarization
fluctuations, roughly an order of magnitude smaller than the temperature
fluctuations \cite{kos99}.

\subsection{Power Spectra}

The temperature fluctuations have been measured and studied in detail
for the past decade, beginning with the watershed COBE detection \cite{gor94},
while
the first measurement of polarization fluctuations by Kovac and collaborators
has been discussed
at this conference. So far, on angular scales down to
about a degree, the temperature fluctuations appear to possess a
Gaussian random distribution \cite{par01,sha02}. To the extent that the
fluctuations are Gaussian, they are described completely by their
power spectra. A parity-invariant distribution of primordial
fluctuations will result in four non-zero power spectra: temperature
$C_l^T$, two polarization $C_l^E$ and $C_l^B$, and the cross-correlation
$C_l^{TE}$ \cite{kam97,zal97c,lue99} (where $l$ is the multipole moment,
inversely proportional to angular scale with $l=200$ corresponding
to one degree). These are the observables we will
primarily consider here, although additional non-Gaussian temperature
structure in the maps will also be very interesting to probe.

Theoretically, we can model how well a given measurement will probe
the power spectrum, given its sky coverage, angular resolution,
and sensitivity \cite{kno95,kam97,zal97c}. Experimentally, 
the MAP satellite will map the
full sky with an angular resolution of around 12 arcminutes and
an effective sensitivity of around 25 $\mu$K, in five frequency bands
ranging from 30 GHz to 150 GHz. A few ground-based experiments have
displayed higher angular resolution and sensitivity over far smaller
regions of the sky; see Max Tegmark's contribution to these proceedings
for a current compilation of power spectrum measurements.

\subsection{Cosmological Parameters}

The intense interest in microwave background temperature fluctuations
has been fueled largely by the realization that the power spectrum
contains much information about cosmological parameters which describe
the fundamental properties of the Universe \cite{jun96,zal97a,bon97}: the
Hubble parameter $h$; the densities of baryons $\Omega_b$, dark
matter $\Omega_{cdm}$, and ``dark energy'' $\Omega_{\Lambda}$; the
primordial power spectrum of scalar perturbations
parameterized by an amplitude $A_s$ and
power law index $n_s$; the same for tensor fluctuations $A_t$ and $n_t$;
and the optical depth to the surface of last scattering $\tau$. 

The ability of the microwave background to constrain these parameters
hinges on the existence of acoustic oscillations of the primordial
plasma at the last scattering surface. These oscillations are phase-coherent
in simple cosmological models, since each $k$-mode has a characteristic
cosmological time at which it enters the horizon, and result in the
well-known series of ``acoustic peaks'' in the power spectrum. The
amplitudes and angular scales of these peaks in turn depend on the
entire set of cosmological parameters. Precision measurements of the 
power spectrum constrain the parameters, with only one essential
degeneracy. The microwave background provides, by far, the single
most powerful set of constraints on the basic properties of the Universe,
and indeed is on the verge of giving definitive answers to most of
the historically most important and vexing questions of classical 
observational cosmology.

\subsection{Parameter Constraints}

Extracting parameter constraints from microwave background
power spectrum measurements is conceptually simple, but somewhat
difficult in practice. Finding the best-fit cosmological model
for a given power spectrum is not hard, but evaluating an
error region requires looking around in a multi-dimensional
parameter space. Approximations based on linear extrapolations
of the likelihood in the above parameters give the right
qualitative answer \cite{jun96}, but are inadequate to support
the currently available data. Brute-force analyses on parameter
space grids have been performed \cite{wan02}, but are not in general
sufficiently accurate or flexible for upcoming data.

Monte Carlo techniques are much better, provided that the
power spectra for a given cosmological model can be evaluated
efficiently enough \cite{chr01,kno01,lew02}. To this end, it is
convenient to use a different set of cosmological parameters 
which better reflect the physical effects determining the
acoustic peak structure in the power spectra. 
The following set of physical parameters has several advantages, including
being largely uncorrelated and having nearly linear power spectrum
dependence \cite{kos02a}: 

\begin{itemize}
\item
${\cal A}\equiv r_s(a_*)/D_A(a_*)$, where $r_s(a_*)$ is the sound horizon
at the time of last scattering, and $D_A(a_*)$ is the angular diameter
distance to the surface of last scattering. This parameter determines
the angular scale of the acoustic peaks.
\item
${\cal B}\equiv \Omega_b h^2$, the baryon density.
\item
${\cal V}\equiv \Omega_\Lambda h^2$, the vacuum energy density.
\item
${\cal R}\equiv a_*\Omega_{\rm mat}/\Omega_{\rm rad}$, the ratio of
matter to radiation energy density at last scattering.
\item
${\cal M}\equiv (\Omega_{\rm mat}^2 + a_*^{-2}\Omega_{\rm rad}^2)^{1/2} h^2$,
which is approximately a degenerate direction in the space of physical
parameters. It is fixed if the number of neutrino species is assumed
known.
\item
${\cal Z} \equiv e^{-2\tau}$, which parameterizes the effect of
reionization.
\item
$n$, the primordial scalar perturbation power spectrum index.
\item
${\cal S}$, the amplitude of the CMB power spectrum due to scalar
perturbations at large $l$ for $n=1$; for fixed ${\cal S}$, varying
${\cal Z}$ changes the power spectrum only at large scales.
\item
${\cal T}$ and $n_T$, the tensor perturbation amplitude and spectral index.
Note it is far more convenient to use ${\cal S}$ and ${\cal T}$ separately,
rather than their sum and ratio as is customary.
\end{itemize}

These parameters allow an extremely efficient evaluation of the power
spectrum for a given set of cosmological parameters over a large
region of parameter space via simple functional approximations; they
make Monte Carlo evaluations of parameter space error regions far less
demanding computationally \cite{kos02a}.  Also, the parameters give
insight into the fundamental physical effects the cosmological parameters
have on the power spectrum. Some
conclusions are immediate; for example, with these parameters, it is
clear that power spectrum measurements at angular scales $l>1000$ will
give very little additional constraint on any physical parameters
besides $n$ and ${\cal S}$, because the change in the power spectrum
for $l>1000$ when the other parameters are varied is negligible.  A
full sky map with MAP's angular resolution and sensitivity will
provide 1-$\sigma$ constraints of approximately 0.5\% for ${\cal A}$,
2\% for ${\cal S}$, 3\% to 5\% for ${\cal B}$, ${\cal R}$, and $n$,
30\% for ${\cal M}$; ${\cal V}$ is essentially undetermined by the CMB
alone \cite{kos02a}.

\section{The Microwave Background and Fundamental Physics}

Beyond simple cosmological parameter estimation, the microwave
background can provide other data with potential impact on fundamental
physics.

\subsection{Primordial Power Spectrum}

Generally, the primordial power spectra of scalar and tensor perturbations
have been parameterized as power laws. This approximation appears to
be fairly good in the case of scalar perturbations, and slow-roll
models of inflation predict power law spectra. However, more complicated
inflation models generically predict departures from exact power laws
\cite{sal89},
and the microwave background fluctuations have considerable power to
measure the primordial power spectrum without prior assumptions about
its shape. 

The angular scales between $l=1000$ and $l=3500$ are largely unaffected
by variations in the physical parameters in the previous section, and
thus reflect the primordial power spectrum directly. However, this is
only a factor of 3 in angular scale. To constrain the primordial perturbations
over a significantly wider range of scales, the region between $l=2$ and
$l=1000$ must be probed, but here the microwave background fluctuations
vary greatly with the other cosmological parameters. The temperature
fluctuations alone exhibit a virtual degeneracy between the primordial
power spectrum and the effect of the other cosmological parameters on
this range of scales \cite{wan99}. The power spectrum in this range cannot
be probed effectively by temperature fluctuations alone,
without further assumptions or other measurements of the
cosmological parameters. However,
the acoustic peaks in the polarization power spectrum are generally out
of phase with those in the temperature power spectrum, which breaks
this degeneracy to a large extent. Accurate measurement of both the
temperature and polarization power spectrum between $l=2$ and $l=3500$
has the potential to measure directly the primordial power spectrum
of scalar perturbations over a significant range in wavenumber \cite{teg02}.

\subsection{Initial Conditions}

Usually, ``adiabatic'' initial conditions are assumed, which means that
the fractional density fluctuations in each particle species are identical.
Again, this is the natural prediction of the simplest inflationary models,
but in general other fluctuations are possible. Efforts have been made
to classify all such ``isocurvature'' fluctuations \cite{buc00}, although
no mathematically rigorous classification has yet been obtained. 
Arbitrary initial conditions greatly expand the parameter space of 
possible models and reduce the ability to determine the cosmological
parameters \cite{buc02,tro01}. 
The current CMB measurements show that the primordial
perturbations are not far from adiabatic; power spectrum measurements of
both temperature and polarization have the potential to put fairly sharp
limits on the contributions from any other isocurvature components, which
in turn could constrain the number of dynamical fields in inflation and
their couplings to each other.

\subsection{Gaussianity}

The statistical distribution of the temperature fluctuations on the
sky, and not just their power spectrum, is another way to probe the
characteristics of the primordial perturbation. Simple inflation
models predict the primordial perturbations should be Gaussian random
distributed; departures from this prediction would signify a more
complicated inflation mechanism or other new physics \cite{ber02}.
Gaussianity is a highly special case of all possible fluctuation
patterns; no single definitive test for non-Gaussianity exists. Various
techniques have been developed in the context of the microwave
background (e.g. \cite{roc01,win97,fer97}), but so far all measurements of
the temperature fluctuations down to sub-degree angular scales show no
statistically significant departure from Gaussianity
\cite{par01,sha02,kun01}.

\subsection{Gravitational Waves}

Inflation generically predicts primordial tensor perturbations, or
gravitational waves, with an amplitude proportional to the energy
scale of inflation. Tensor perturbations can be cleanly separated from
scalar perturbations by observing a B-polarization signal, which is
produced by tensor but not by scalar perturbations
\cite{kam97b,kam98,zal97c}.  Current measurements of the temperature
power spectrum limit the amplitude of tensor perturbations to be no
larger than around 20\% of the scalar perturbation amplitude. 
Recently, Knox and collaborators have shown that the B-polarization
induced by gravitational lensing provides a lower limit to the
amplitude of tensor perturbations which can be detected via microwave
background polarization, corresponding to an inflation energy scale of
around $3\times 10^{15}$ GeV \cite{kno02}.  If inflation occurred at
energy scales higher than this, we can eventually expect direct
confirmation of the inflationary scenario via detection of the
microwave background polarization signal produced by the inevitable
inflationary gravitational waves. Note this energy scale is generally
below the coupling-constant unification scale in 
GUT models of particle physics.

\subsection{Topology}

A novel application of full-sky microwave background maps is a strong
test for non-trivial large-scale topology of the Universe.  While the
assumption of local homogeneity and isotropy determines the Friedmann
equation governing the expansion of the Universe, it says nothing
about large-scale topology (see \cite{lac95} for a review).
%Any topological structure
%which is larger than the current horizon scale is obviously
%unobservable. However, if topology on scales smaller than the horizon
%exists, then certain regions of the Universe will be observable
%along more than one geodesic, at different ages and in different
%sky directions. Direct topological tests looking for multiple images
%of, say, individual galaxies in different parts of the sky are quite 
%difficult due to evolution effects. Such analyses can clearly place
%some lower limits on the topology scale, but are largely inconclusive
%on cosmological scales \cite{lac95}.
%
Cornish, Spergel, and Starkman \cite{cor98} have proposed an elegant
topological test using the cosmic microwave background.  They noticed
that since the last scattering surface is a sphere, any
non-trivial topology, which translates into intersecting the last
scattering surface with multiple copies of itself, will lead to
circles on the sky with identical temperature patterns. (This is
actually only approximately true, because part of the observed
temperature fluctuation comes from Doppler shifts at the last
scattering surface instead of temperature fluctuations.) The number of
these circles and their relative directions and orientations can be
used to reconstruct the topology of the Universe \cite{wee98}. Only a small
number of topologies are possible in Universes with positive or zero
spatial curvature, but negatively curved ones can support a rich
variety of topologies, including some relatively small compared to the
curvature scale \cite{hod94}.  The MAP satellite has sufficient
resolution to perform this test.  Related ideas have also been studied
\cite{lev98,bon00}; see \cite{lev02} for a recent review.

\subsection{What Cosmology Offers High-Energy Theorists}

Here is a brief list of cosmological information that is relevant to
fundamental physics, probed mostly by the microwave background:
%in the sense of constructing models of the
%basic building blocks of matter (i.e. particles, strings) and their
%interactions. This list assumes that our current standard cosmological
%model is basically correct. 
%most of these can be probed only via the microwave background.

\begin{itemize}
\item
The primordial power spectrum of fluctuations over three decades in
wavelength. 
\item
Limits on or characterization of non-Gaussianity in the primordial fluctuations.
\item
Limits on or characterization of any isocurvature components in the
primordial fluctuations.
\item
Detection of primordial gravitational waves, or a limit on their amplitude.
\item
Limits on or detection of global topology.
\item
Expansion rate of the Universe at the epoch of nucleosynthesis 
\cite{carroll02}.
\item
Expansion rate of the Universe at recent epochs, either from
direct observation of standard candles like SNIa \cite{per99,gar98} or from
nonlinear fluctuations in the microwave background (see below).
\item
Constraints on dark matter properties. These come from galaxies and clusters,
but are less clean than microwave background conclusions, and are currently
in an unsettled state (see, e.g., \cite{deb01}).
\item
Direct detection of gravitational radiation from the early Universe. 
This is the other potential direct source of information about the 
early Universe besides the microwave background, and could potentially
probe the electroweak phase transition and similar epochs \cite{kam94,kos02b}.
\end{itemize}

This is the ground on which high-energy theorists must meet cosmologists.
A general and largely unaddressed question is in what ways these
sources of information can constrain fundamental theories of matter and its
interactions.  Optimistically, eventually we will have
a candidate ``theory of everything'' with unavoidable
cosmological predictions, and the above data sources will serve as a
strong test of any such theory.

\section{Small-Scale Nonlinear Fluctuations}

At small angular scales, $l>3500$, the power spectrum of microwave
background fluctuations becomes dominated by secondary effects from
nonlinear structures at recent epochs, not by the primary fluctuations
from linear perturbations in the early Universe. MAP will produce the
definitive measurement of microwave background temperature fluctuations
out to $l=800$, and the upcoming Planck satellite will measure
the temperature (and polarization) fluctuations out to $l=3000$
in many frequency bands before the end of this decade. At that point, 
observations of the primary temperature anisotropies will 
largely be exhausted. 
Presently, attention is shifting to the small-scale, non-linear
fluctuations, and particularly the fluctuations induced by clusters
of galaxies, the largest gravitationally bound objects in the Universe.
Clusters are potentially powerful tracers of the growth of cosmic
structure, and thus have the potential for further probes of the
fundamental properties of the Universe, though these conclusions
will necessarily be accompanied by more severe systematic challenges
than for probes based on the primary, linear CMB fluctuations.
Below, several interesting aspects of nonlinear CMB fluctuations are
sketched.

\subsection{Thermal Sunyaev-Zeldovich Effect}

The largest non-linear signal is that of the thermal Sunyaev-Zeldovich
effect \cite{bir99,car02}, the spectral distortion that occurs when
microwave background photons, with an initially blackbody spectrum,
are Compton scattered by hot electrons, generally in clusters
of galaxies. Lower energy photons are boosted to higher energies;
the spectrum amplitude at the ``null'' of the effect, around 218 GHz (depending
slightly on the density and temperature of the electrons), remains
constant. In the direction of galaxy clusters, therefore, the microwave
radiation appears cooler for frequencies below the null, hotter for
frequencies above. The amplitude of this effect can be as large
as 1 mK for large clusters of galaxies, much larger than
the primary temperature fluctuations. The total
spectral distortion is proportional to the product of the electron
density and the electron temperature, integrated along the line of
sight. 

The thermal SZ effect has now been detected for numerous clusters of
galaxies, at roughly arcminute angular resolution
(e.g. \cite{mau00,ree02}). Besides providing direct information about
the density and temperature of the gas in galaxy clusters, the great
utility of the SZ effect is its independence of cluster redshift:
since the induced spectral distortion remains as the microwave
background radiation propagates, a given galaxy cluster will produce
the same observed SZ distortion independent of its distance from the
observer. This is in marked contrast to other methods of observing
clusters from their direct emission of radiation.  SZ observations
hold the promise of cluster catalogs with relatively simple and
complete selection functions, which in turn are necessary for any use
of galaxy clusters as precision cosmological probes.

\subsection{Kinematic Sunyaev-Zeldovich Effect}

A related but smaller nonlinear effect results from the Doppler shift
experienced by photons scattering from electrons moving with a
coherent peculiar velocity. This bulk velocity induces a blackbody temperature
shift of the microwave photons proportional to the radial component of
the electrons' peculiar velocity.  (In the mildly nonlinear regime of
structure, this effect is known as the Ostriker-Vishniac effect
\cite{jaf98}.)  In galaxy clusters, the typical amplitude of this
temperature shift is a few $\mu$K, much smaller than the thermal
effect. The two can be separated by their spectral dependences, in
principle. Even though galaxy clusters are highly dynamic objects with
significant internal bulk flows due to mergers, the average kinematic
SZ signal provides a largely unbiased measure of the cluster's
peculiar radial velocity \cite{nag03}.  If the kinematic SZ
distortion can be extracted reliably for individual galaxy clusters,
then galaxy clusters can be used as tracers of the cosmic peculiar
velocity field out to redshifts beyond $z=1$. Current peculiar
velocity surveys extract velocities by estimating the distance to
galaxies and then subtracting the inferred Hubble velocity from the
observed redshift velocity to obtain a peculiar velocity
\cite{wil97}. Such a procedure quickly becomes dominated by
systematic errors at cosmologically modest distances due to the difficulty of
accurate distance estimation.  The great advantage of the kinematic SZ
effect, in comparison, is that it provides a direct peculiar velocity
estimate without requiring a distance determination. A peculiar
velocity map over a substantial portion of the observable Universe
will provide a sharp test of the gravitational instability paradigm of
structure growth \cite{jus99} and a strong consistency check with
surveys of the cosmic density field \cite{dod02,per01}.

\subsection{Weak Gravitational Lensing}

As the microwave background radiation propagates from the last
scattering surface to the observer, its geodesics will be altered by
the presence of intervening matter inhomogeneities. This gravitational
lensing has only small effects on the power spectrum of the microwave
background fluctuations \cite{sel96b}, but does change the
pattern of the radiation, inducing a specific form of non-Gaussianity
\cite{ber97}. Lensing creates a correlation between two-point
correlations on degree scales and four-point correlations on smaller
scales.  Recently, algorithms have been developed to reconstruct the
lensing mass distribution given a lensed temperature map. Temperature
information alone can give the correct qualitative structure, while a
high-sensitivity polarization map on scales of a few arcminutes can
determine the projected lensing mass distribution to good accuracy on
sub-degree scales \cite{hu02}. The lensing mass distribution can
also be determined from shear measurements of background galaxies on
comparable angular scales, providing a valuable cross-check. Microwave
background lensing has the advantage of a well-defined source
redshift, along with completely different systematic errors for these
challenging observations.

\subsection{Strong Gravitational Lensing}

In the sky regions of galaxy clusters, the large mass concentration
significantly distorts the microwave background temperature pattern.
This strong gravitational lensing signal has several distinct features.
Most notably, it produces a double-lobe distortion aligned with the
temperature gradient of the background fluctuations \cite{sel00}.
This distortion can be used to reconstruct the cluster mass profile,
in principle: while observations of background galaxy shapes yield
only information about the relative shear field of the mass distribution,
lensing of the microwave background provides direct information
about the displacement field induced by the mass distribution. The
characteristic angular scale of the strong lensing distortion is
an arcminute; on this scale, the primary CMB fluctuations have essentially
no power. Thus, in certain regions of the sky where the primary fluctuations
are especially regular, the lensing displacement field from a cluster
can likely be modelled with good precision and used to estimate
cluster masses. Since the displacement is estimated directly, this
method has no mass sheet degeneracy, like optical galaxy lensing estimates
of cluster masses, and has no uncertainties related to the background
source redshift. Ultimately, the accuracy of this cluster mass determination
method will depend on how well the background microwave temperature
distribution can be modelled, and on how well the blackbody lensing distortion
can be separated from the blackbody kinematic SZ distortion. Accurate
cluster mass determination is crucial for directly measuring $N(M,z)$, the
number density of clusters at a given mass and redshift. This
function is a highly sensitive probe of the recent growth of structure
and constrains the cosmological constant or other ``dark energy'' 
contributions, as well as small neutrino masses.

\section{Future Experimental Prospects: ACT}

Experimental techniques for measuring the microwave background
radiation are, remarkably, advancing at an accelerating pace.  The
various kinds of observations outlined in the previous section
require, roughly, arcminute resolution maps with micro-Kelvin
temperature sensitivity. Such measurements are on the menu for the
coming half-decade. As an example, I provide a brief overview of the
Atacama Cosmology Telescope (ACT), an experimental collaboration
between Princeton, U.~Pennsylvania, Rutgers, NASA Goddard, NIST, and
several other smaller partners. This collaboration exhibits many
technological and organizational characteristics which we anticipate
will come to dominate the microwave background field over the coming
decade.  Experiments of comparable scope and ambition are also being
planned by other groups.

The ACT collaboration plans to construct a custom-designed 6-meter
off-axis telescope which is optimized to minimize the systematic
errors which can easily dominate any precision measurement of the
microwave background radiation. It will scan the sky at fixed
elevation by rotating on a turntable; the entire telescope, including
ground screens, will rotate as a unit to eliminate any change in
side lobes. Constant-elevation scans greatly aid 
in controlling systematic errors, since the
atmospheric emission changes by many hundreds of $\mu$K for elevation
changes smaller than a degree. A similar design philosophy was used
successfully in the Saskatoon \cite{wol93} and MAT \cite{mil02}
experiments. The telescope will be cited on Cerro Toco in the Atacama
Desert of the Chilean Andes, near the site for the ALMA
interferometer. Site studies show that the atmospheric signal at
microwave frequencies will generally be smaller than our detector
noise for all scales $l>100$.  At this latitude, scans at a constant
45$^\circ$ elevation with an amplidude of $\pm 1.5^\circ$
can be combined over 24 hours to produce a map of
an annular region of sky approximately 2$^\circ$ wide by 120$^\circ$
degrees around. About half the scan will cover the galactic plane and
will be useful for galactic astronomy; the other half will be of
cosmological quality.

ACT's novel detector technology will enable it to produce maps of
unprecedented sensitivity and angular resolution. We plan to build a
``camera'' composed of bolometer arrays at three different
frequencies, each array containing 1024 square-millimeter transition
edge sensor detectors \cite{irw96,lee96}.
These bolometers are very fast, enabling rapid scanning of the sky.
Each individual bolometer's sensitivity 
is within a factor of two of the bolometers
planned for the Planck satellite; packing them into a large array
gives ACT the raw detector sensitivity required for achieving a
nominal angular resolution of 1.7' (at 150 GHz) with a nominal
sensitivity per pixel of near 1 $\mu$K for a three-month observing
campaign. The largest technological challenge of the experiment is
fabricating the detector arrays with their associated SQUID
multiplexors \cite{che99} and other
back-end electronics. Current state-of-the-art bolometer arrays employ
tens of bolometers; we aim to increase this by a factor of a
hundred. The anticipated time scale for building the experiment and
observing for two seasons is five years.

The microwave observations from ACT will be combined with an integrated
optical and X-ray observing campaign aimed at the galaxy clusters
discovered via their SZ signals. Optical observations with the
Southern African Large Telescope (currently under construction) and
other large telescopes will provide redshifts and galaxy velocity
dispersions for hundreds of clusters out to redshift unity, while
X-ray observations will give information about gas temperature and
density. Through this combination of observations, we aim to 
construct the most complete and useful cosmological cluster sample
to date. We will begin to probe all of the small-scale
signals outlined in the previous section. 

As spectacular as
the advances in microwave background physics have been over the
past decade, we can reasonably expect another decade of comparable
discovery, with remarkable implications for cosmology, fundamental
physics, and astrophysics.

\begin{theacknowledgments}
At the time of this writing, 
the Atacama Cosmology Telescope is a currently pending NSF
proposal; Lyman Page (Princeton University) is the Principal
Investigator. The author thanks all of the collaborators on this
proposal, who are too numerous to list here.  This work has been
supported in part by NASA's Space Astrophysics Research and Analysis
Program, and by the Cottrell Scholar program of the Research Corporation.
\end{theacknowledgments}

\bibliographystyle{aipproc}

\bibliography{cmb}

\end{document}